\documentstyle[preprint,aps,prl,epsfig]{revtex}
\begin{document}
\draft
\title{Structure of Electrorheological Fluids}
\author{U.Dassanayake and S.Fraden}
\address{Complex Fluids Group,~Martin Fisher School of Physics,~Brandeis University,~Waltham, MA 02454}
\author{A. van Blaaderen}
\address{Van 't Hoff Laboratory, Debye Inst., Utrecht University, The Netherlands  and FOM Inst. for Atomic and Molecular Physics, Amsterdam, The Netherlands}
\date{\today}
\maketitle

\begin{abstract}

Specially synthesized silica colloidal spheres with fluorescent cores were used
as model electrorheological fluids to experimentally explore structure formation
and evolution under conditions of no shear. Using Confocal Scanning Laser
Microscopy we measured the location of each colloid in three dimensions. We
observed an equilibrium body-centered tetragonal phase and several
non-equilibrium structures such as sheet-like labyrinths and isolated chains of
colloids. The formation of non-equilibrium structures was studied as a function
of the volume fraction, electric field strength and starting configuration of
the colloid. We compare our observations to previous experiments, simulations
and calculations.

\end{abstract}
\newpage

\bf  INTRODUCTION \rm

Electrorheological (ER) fluids are suspensions of dielectric particles (usually
of size 1-100$\mu$m) in non-conducting or weakly conducting solvents. For
particles with radii below several $\mu$m Brownian motion is still important and
such dispersions are called colloidal. When electric fields are applied across
these suspensions they tend to show altered viscous behaviour above a critical
value of the electric field, with the apparent viscosities increasing by several
orders of magnitude at low shear rates. Above this critical electric field, at
low shear stresses the suspensions behave like solids, and at stresses greater
than a `yield stress' the suspensions flow with enhanced viscosity. The
rheological response is observed to occur in milliseconds, and is reversible.
This combination of electrical and rheological properties has led to many
proposals for applications of ER fluids in such devices as hydraulic valves,
clutches, brakes, and recently in photonic devices~\cite{Sta83,Alf1998,She99}.

Application of an electric field results in structural transitions in the
colloidal suspension because the interparticle electrostatic interactions due to
polarization are stronger than Brownian forces. The tendency of particles in
suspension to form structures such as chains upon application of an electric
field was reported centuries ago by scientists such as Franklin and
Priestly~\cite{Pri69,Poh78}. Quantitative experiments on the electrorheological
effect were first performed by Winslow in 1949, when he reported that
suspensions of silica gel particles in low-viscosity oils tend to fibrillate
upon application of electric fields, with fibers forming parallel to the
field~\cite{Win49}. Winslow reported that at fields larger than $\approx$3kV/mm
the suspensions behaved like a solid, which flowed like a viscous fluid above a
yield stress that was proportional to the square of the applied electric field.

Particle association is the main cause of the altered rheological behaviour of
ER fluids. The nature of the field induced structures are an important factor in
determining the yield stress and flow behaviour. Moreover, theoretical studies
of the rheological properties of ER fluids are typically performed by subjecting
possible field induced structures to shear stresses and obtaining stress-strain
relationships. Therefore it is important to experimentally study the nature of
the particle aggregates quantitatively in real-space. In this paper we describe
an experimental study of structure formation in a model ER fluid in the absence
of shear fields, by direct visualization of the fluid.

A recent and comprehensive survey of ER fluids, where the issue of particle
aggregation is addressed, is provided by Parthasarathy and
Klingenberg~\cite{Par96}. Tao and Sun have predicted the (zero-temperature)
ground state for ER fluids to be a body centered tetragonal (BCT) structure (see
figure 1). Their result was obtained for suspensions of uniform spheres by
treating the ER fluid as a suspension of point dipoles in a dielectric fluid and
by only taking energy considerations (neglecting entropy considerations) into
account~\cite{Tao89}. An experimental verification of this proposed ground state
structure has been provided by laser diffraction studies of ER systems
consisting of 20$\mu$m diameter glass spheres suspended in silicone oil, where
entropy considerations are indeed not important~\cite{Che92}. Halsey and
Toor~\cite{Hal90} have described the evolution of structure as occurring in two
identifiable stages. The particles first chain along the electric field, and
then aggregate into dense structures that take the form of columns aligned with
the electric field. In recent computer simulation studies, Martin, Anderson and
Tigges have observed the formation of two-dimensional `sheets' of particles
(described later) as an intermediate state in structure
evolution~\cite{Mar98,Mar99}. In their simulations, Martin and coworkers
describe the evolution of structure in an ER fluid consisting of 10,000
particles over a concentration range of 10-50\% volume fraction. They study the
mechanics of coarsening and the emergence of crystallinity, and use various
methods of characterizing the structures that evolve, including pair correlation
functions, microcrystallinity and coordination number. Several aspects of their
simulation results are evocative of our observations, but there are several
notable exceptions as described below.\vspace{6mm}

\bf  METHODS \rm\vspace{1mm}

The model ER fluid used in our experiments was a solution of monodisperse silica
spheres (0.525$\mu$m radius, polydispersity in size 1.8\%) in a mixture of water
and glycerol. The spheres were charged and this prevented irreversible
aggregation. The spheres had fluorescent cores of 0.4$\mu$m diameter and were
labeled by fluorescein isothiocyanate (FITC). The colloids were the same as used
in previous work in which particle coordinates were obtained in bulk samples
~\cite{Alf97,Alf95}. The non-fluorescent layer surrounding the core made it
possible to distinguish individual particles even when they were in contact with
each other. The synthesis of this kind of spheres is described in
ref.~\cite{Alf92}. Interparticle attraction was reduced by matching the
refractive index of the solvent to that of the particles, minimizing van der
Waals forces. The matching of refractive indices also enabled visualization in
the bulk of concentrated samples by reducing the multiple scattering of light.
The silica spheres had a refractive index of 1.45, and index matching was
achieved by using a solvent mixture of 16 w.\% water (refractive index 1.33) and
84 w.\% glycerol (refractive index 1.48). The viscosity of such a mixture (84.4
cP) also allows a larger time window in which the dynamics of structure
formation can be followed with confocal microscopy. We estimate that the silica
spheres had a dielectric constant of 3.7 (at 500kHz) while the solvent had a
dielectric constant of $\approx$49 (at 500kHz), values which are close to the
values at 0Hz. While the refractive indices were approximately matched at light
frequencies, we could still obtain a dielectric mismatch between the spheres and
the solvent at frequencies in the 500kHz range in which we applied AC electric
fields. The ER fluid was placed in a cell, which consisted of parallel
microscope slides that were coated with indium tin oxide (ITO), a transparent
conductor. The ITO coating had an electrical resistance of 100 $\Omega$ per
square inch. One slide was $\approx$1.2mm thick, while the other was
$\approx$150$\mu$m thick, and served as a cover slip. An insulating sheet of
Kapton (DuPont) was placed between the electrodes, serving as a means for
adjusting the electrode gap (between 5-100$\mu$m). The ER fluid was placed in a
$\approx$3mm diameter hole cut out of the Kapton sheet, which served to confine
the fluid, and the electrodes were aligned horizontally.

The electrodes were connected to a power supply providing a uniform electric
field. Electric fields of strength $\approx$1kV/mm and frequency $\approx$500kHz
were used. A Krohn-Hite (Avon, MA) Model 7602M wideband power amplifier and a
waveform generator from Wavetek (San Diego, CA) were used to produce the
electric fields. The electric field strength was of the same order of magnitude
as that used in previous experiments on ER fluids. The frequency was chosen so
that effects due to the polarization of the double layer could be neglected, and
the polarization of the spheres could be attributed to the dielectric mismatch
between the spheres and the solvent.

The observations were made using confocal scanning laser microscopy
(CSLM)~\cite{Wil95} which is a technique that images individual planes in a
sample by rejecting most of the fluorescent light emitted by particles out of
the imaging plane. Additionally, confocal microscopy yields better resolution
than conventional light microscopy both along the optical axis and in the image
plane. The resolution produced is $\approx 0.6 \mu$m along the optical axis, and
$\approx 0.2 \mu$m in the image plane. In our experiments we used CSLM to obtain
sequences of digitized 2-dimensional images of the sample at planes separated by
$\approx 0.1 \mu$m. These images were then computer-analyzed to obtain the
3-dimensional coordinates of the particles present in the sample. These
coordinates could then be used to analyze the structure, and to render the
sample as a 3-dimensional graphical object. Typically it took several minutes to
obtain a three dimensional data set of 512x512x100 voxels (e.g., $40
\mu\mbox{m} \times 40 \mu\mbox{m} \times 20 \mu$m) and several seconds to
obtain a single 2D image plane of 1024x1024 pixels (e.g., $40 \mu\mbox{m} \times
40 \mu\mbox{m}$) either as a plane perpendicular or parallel to the optic axis.
The experiments were performed on a system consisting of a Leica inverted
microscope with a 100x1.4NA oil lens with a Leica TCS confocal attachment.

The process of obtaining the 3-dimensional coordinates of the particles from the
confocal microscope images was similar to that described in ~\cite{Alf97,Alf95},
and is illustrated schematically in figure 2. The stacks of images were of
sequential planes perpendicular to the optical axis, about 0.1$\mu$m apart. The
(spherical) particles imaged through the confocal system appear as circular
regions in the digitized images (see figures 3,4,5) because of the circular
symmetry of the confocal microscope point spread function(psf) in the plane
perpendicular to the optical axis (the xy plane). Because of the finite extent
of the psf along the optical axis (z axis), a given particle is imaged in
several consecutive image planes as circular regions of varying intensity and
size. Each image was first analyzed to find the centers of each region present.
This was achieved by identifying each region above a chosen threshold of
intensity as resulting from an individual particle, and identifying the
intensity-weighted average position (xy coordinate) of each region as the center
(care was taken to identify overlapping particle regions). Neighbouring planes
were then looked at, and region centers with approximately the same xy
coordinates were identified as belonging to a single particle, forming a
`string' of centers for a single particle. The final xyz coordinates of each
particle was obtained by finding the center of intensity along the `strings'.
Since there is a distribution in the sizes of the cores there is also a
distribution in the fluorescent intensities of the particles. Furthermore, the
fluorescence photobleaches, and the detected intensities diminish in the bulk of
the sample due to optical abberations. Therefore the algorithm had to be applied
at many iterations of the threshold value. The initial threshold value was set
at the intensity of the most weakly fluorescing particles. These were then
identified as distinct particles using the above described algorithm, while
particles of higher intensity blended together. Once the weakest particles were
identified, they were removed from the dataset, the threshold was increased, and
the search for particles was repeated. This was repeated at several iterations
of the threshold value. The particle coordinates found were verified by marking
the calculated coordinates in the raw data and visually inspecting the results.
The data was rendered in 3 dimensions on a Silicon Graphics Indigo$^{\rm{TM}}$
platform using programs implementing native graphics library routines. Data was
also visualized through web-based 3d browsers using the VRML format (specified
at http://www.vrml.org). IDL, a programming environment useful for visual data
analysis (from Research Systems Inc., Boulder, CO), was used extensively for the
manipulation of images.\vspace{6mm}

\bf  OBSERVATIONS \rm\vspace{1mm}

We found the structure formation to proceed through a sequence of nonequilibrium
structures depending on the initial conditions of the suspension and the
strength of the applied field. The structure formation occurred most rapidly at
early times, within a few seconds. We found it most useful to describe the
structure development according to the volume fraction of spheres used. The
following is a summary of our observations.

At low fields, below $\approx$100V/mm, where the interparticle electrostatic
interaction energies were low compared to thermal energies, no significant
particle association was observed. The spheres tended to sediment to the bottom
electrode, having a density larger than the glycerol-water solvent. The
structural observations were performed at field strengths of $\approx$1000V/mm
where field induced structures, such as chains of touching particles, that
formed were not observed to break up due to thermal fluctuations, implying that
the electrostatic energy at contact was many kT. However, we observed
significant Brownian motion even in the final crystalline states. A particle in
a BCT crystal with an applied field of 0.5kV/mm typically had a measured mean
square displacement (in the plane perpendicular to the electric field) of about
5\% of the lattice spacing.

At the lowest observed particle volume fractions of about 10\%, and with the
particles initially distributed throughout the volume of the solvent, rapid
application of fields of strength $\approx$1000V/mm resulted in the formation of
field-aligned chains and `sheets'. Chains are linear associates of touching
spheres aligned along the field, with a wide variation in length. Sheets are
hexagonally ordered, 2-dimensional structures, aligned with the field direction
(figure 3). At this particle concentration the number of chains formed initially
was larger than the number of sheets, and as the concentration was increased,
the presence of chains decreased relative to the presence of sheets. Sheets
appeared within seconds of the field being turned on, and rapidly formed into a
complex, interconnected labyrinthine formation (figure 4a). They formed
initially at the electrodes and grew away from both electrodes towards the
middle of the electrode gap. The chains appeared initially throughout the
sample, with more in the middle of the electrode gap where they rapidly
transformed into BCT structures by attracting each other. We determined the
structures to be BCT by direct measurement of the crystal dimensions (figures
5,6). The sheets transformed into BCT structures over a time of a few hours by
annealing together, beginning in the regions away from the electrodes, and
growing towards the electrodes. The BCT structures that formed rearranged
themselves into a network of misaligned BCT regions such as seen in figure 4b
and only coarsened very slowly over the time of observation of 1-2 days by
collective motions. The first 2-3 layers of spheres at the electrodes remained
in hexagonal planes parallel to the electrodes throughout the observation
(figure 6). An explanation may be that the spheres at the electrodes are
strongly attracted to their image dipoles created by the conducting electrodes.
We did not have enough data of particle coordinates to attempt to characterize
the structure in terms of a local order parameter and thereby to quantify the
structure formation over time~\cite{Mar98}, although in principle this is
possible to do.

Table 1 shows calculated values of the dipolar energy per particle at large
electrode gaps for various particle arrangements (from refs~\cite{Tao89}
and~\cite{Mar98}). The energy per particle is in units of
$p^2$$/$$a^3$$\epsilon_f$, where
$p=a^3$$\epsilon_f$E($\epsilon_p$-$\epsilon_f$)/($\epsilon_p$+2$\epsilon_f$).
$\epsilon_p$ and $\epsilon_f$ are the dielectric constants of the particle and
solvent respectively, and E is the magnitude of the electric field. The BCT
structure is the most favourable structure, while sheets have a value in between
that of chains and BCT. The value given for sheets is that for large hexagonally
packed sheets and is independent of the arrangement of the spheres within the
sheets. Neglecting the presence of the electrodes, because of the symmetry, the
dipolar energy per particle in a hexagonally packed planar field-aligned sheet
is independent of its orientation~\cite{Tao96}. However, in our observations we
saw a distinctive orientation for the sheets. As seen in figure 3, the sheets we
observed can be considered to be composed of strings of nearest neighbours of
spheres that are tilted by 30$^\circ$ with respect to the electric field axis
(as opposed to our expectations of observing hexagonal packing created by a
series of offset sphere chains aligned with the field). This observed
orientation for the sheets is probably created by the presence of the
electrodes. Because of the strong attraction between spheres and their image
dipoles at the electrodes, a layer of spheres forms on the electrode, which
nucleates the growth of hexagonal sheets having the configuration observed.
Figure 7 shows a calculation of the energy per dipole (in units of
$p^2$$/$$a^3$$\epsilon_f$) for various structures where the calculations include
the presence of the electrodes by including interactions with image dipoles. The
calculations assume fixed identical dipoles interacting with the external field
and each other. For each structure considered, the value given is that for a
dipole located in the center of the structure. Two configurations of sheets are
considered, the observed configuration and that formed by a close packed
arrangement of chains. For electrode gaps smaller than $\approx$10 sphere
diameters the observed sheet structure shows the lowest energy. At larger
electrode gaps the BCT structure has the lowest energy, while the difference
between the values for the two sheet configurations decreases. Our experiments
agree with this simple calculation which neglects entropy contributions. For
electrode gaps less than $\approx$12 sphere diameters we observed that the
sheets (and a smaller population of chains) that formed upon application of the
field persisted over the duration of observation (2 days) and were the dominant
structure present, with a small fraction of spheres forming into BCT crystals.
Thus experimentally the sheets seemed to be the equilibrium structures at these
electrode gaps. It is interesting to note that while it would seem natural for
chains of particles to aggregate together to form hexagonal sheets, we did not
observe this in our experiments. Thus the small energy gain (due to the presence
of the electrodes) in forming our observed sheets, while decreasing with
increasing electrode gap, was most likely sufficient to select a preferred
orientation for the sheets, even in very thick samples.

As the particle volume fractions in our samples were increased from 10\% to
15\%, the presence of chains decreased, and increasing proportions of the sample
adopted the metastable sheet state. The time for the sheets to transform into
BCT structures increased as the applied E-field value was increased. We did not
have enough time resolution  to observe the mechanics of the formation of
sheets, which formed within seconds. When the E-field was turned off, the
structures disassociated, driven by Brownian motion, and the spheres returned to
a dispersed state.

At sphere volume fractions $\approx$25\% the structure formation was
investigated under two different starting conditions. In one case, the particles
were allowed to sediment under gravity (figure 8a) in the absence of a field.
The bottom layers of the sediment were observed to be randomly stacked hexagonal
close-packed planes~\cite{Pus89,Ell97}, and the upper layers of spheres were in
a fluid-like state. When an electric field ($\approx$1kV/mm, 500kHz) was applied
across the electrode gap of $\approx$70$\mu$m, the spheres at the interface of
the sediment and solvent began to form field aligned chains that eventually
reached the upper electrode. This occurred within a few (2-3) minutes. Over the
next several hours the spheres in the entire sediment rearranged into field
aligned chains that attracted each other to form columns that spanned the
electrodes (figure 8b). The sphere arrangement within the columns was a BCT
structure. The columns themselves were bridged together by domains of BCT
crystals. After coarsening for a few hours further development of the
column-structures stopped and no further evolution occurred over a period of 2
days. Lowering the electric field slowed down the structure formation. We did
not see sheet-like structures under these initial conditions.

When the electric field was applied across a solution with volume fraction
$\approx$30\%, but with the spheres initially dispersed through the solvent, we
observed a pattern of structure formation similar to that seen at lower
concentrations with similar starting configurations, where labyrinths of small
sheet-like structures, along with isolated chains of spheres developed within
seconds. The sheets evolved into a collection of small sections of
interconnected BCT structures that retained the labyrinthine appearance of the
sheets. When the field was turned off, the structures disappeared within
seconds, as was the case in all our observations.

Observations were made at higher concentrations, volume fraction $\approx$45\%
where the spheres crystallize. In the absence of electric fields the spheres
were arranged in FCC stacked hexagonal layers parallel to the electrodes, with
the top few layers being liquid-like~\cite{Pus89,Ell97}. As is the case in our
observations, it has been observed in other experiments that, except for the
hard-sphere limiting case (thin steric stabilizer layer or very thin double
layer with 0.1 M salt), the crystal structure formed upon sedimentation is
FCC~\cite{Alf99}. When an electric field $\approx$1kV/mm and 500kHz was applied
across the electrode gap ($\approx$80$\mu$m), defects appeared in the hexagonal
structure, and areas of BCT formed in the bulk of the sample over the first
$\approx$10 minutes. After a few hours it was observed that sections of the
colloidal crystal had transformed into BCT order in the bulk of the sample, with
the bottom 5-6 layers remaining hexagonal, and the top 2-3 layers remaining
disordered. There were no sheets observed at this concentration. Figure 9 shows
the transition from hexagonal ordering in the absence of a field, to a mix of
BCT and hexagonal order after the field was on for $\approx$6 hours. It should
be noted that, as indicated in the upper portion of Fig.9b that there is free
space between BCT crystals of different orientation. When the field was switched
off the BCT crystals stayed in the same symmetry but expanded to become 100
oriented FCC crystals, as opposed to the initial 111 FCC symmetry seen before
the field was turned on. When the field is switched on again the crystal goes
through a martensitic transition back to BCT. This ability to tune the crystal
structure by using an electric field could have applications in the field of
photonic crystals.
\vspace{6mm}

\bf DISCUSSION \rm\vspace{1mm}

It is interesting to consider other studies that address the issue of structure
formation. Labyrinthine structures similar to sheets have been observed in
ferrofluids, but the structure within the sheets was not
ascertained~\cite{Law94}. The presence of hexagonal sheets has not been reported
in previous experimental observations on ER fluids, but has been observed in
simulations~\cite{Mar98,Mar99,Mel93,Mel92,Has93}. The simulations of Martin,
Anderson and Tigges~\cite{Mar98,Mar99} show that for sphere concentrations less
than $\approx$30\% a sudden application of the field first induces formation of
short chains parallel to the field.  The chains then attract each other, forming
sheets, which often are bent into tube and spiral-like forms, or form thick
walls . This was in contrast to our observations, where the sheets remained
two-dimensional for long periods of time. Very few BCT domains were found in the
simulations that neglected thermal motion~\cite{Mar98}, probably because the
sheets were prevented from annealing into the ground state BCT crystal.  This is
in contrast to our experiments, where large BCT crystals formed after long
exposures to the field. In their simulations that considered thermal
effects~\cite{Mar99}, Martin, Anderson and Tigges observed increased order,
crystallinity, and larger domain sizes compared to their simulations neglecting
thermal motion~\cite{Mar98}. However a sheet to BCT transition was not seen,
possibly because of the insufficient duration of the simulation. The structure
evolution we observed followed the general pattern of the
simulations~\cite{Mar98,Mar99}, although we did not observe the sheets to form
from the association of chains. However, the time resolution of our observations
was not sufficient to study the initial association in more detail.

Halsey and Toor~\cite{Hal90,Too93} studied structure evolution by considering
that the shape of a particle aggregate would be a droplet, modeled as a prolate
spheroid. The droplet is assumed to grow with time, in a quasi-equilibrium
manner, as individual particles attach to it. It elongates towards the
electrodes as it grows, with its shape being determined by balancing the bulk
and surface electrostatic energies due to the dipoles it contains, forming into
a column spanning the electrodes. Individual columns will aggregate over time
towards a bulk phase segregation. This model assumes the concentration of
particles is low, and that the droplet is always in equilibrium, which requires
that the electric field is low compared to thermal energies. They point out that
the equilibrium droplet model fails at high fields, where columns will form
rapidly in a non-equilibrium manner but growth will become arrested before
equilibrium bulk phase separation occurs. We did not observe droplet-like
structures. At the volume fractions we used, and with the rapidly quenched
relatively high electric fields we used, structure formation did not occur
particle by particle in a quasi-equilibrium manner as in the droplet model, but
rather as a association of chains and sheets. We observed column formation at
volume fractions of $\approx$25\% and with the colloid initially sedimented, but
the columns did not coarsen to allow bulk phase separation over a period of
observation of 2 days. Martin and colleagues have performed two-dimensional
light scattering studies on a model ER fluid using particles and field strengths
similar to those used by us~\cite{Mar98a}. They observed a two-stage, chain to
column structure formation under conditions of no shear. Although sheet
formation might be expected under the conditions used, they did not observe
sheets. This may be because light scattering methods alone are insufficient to
detect and interpret sheet-like structures.

Melrose~\cite{Mel93,Mel92} has performed Brownian dynamics simulations on ER
fluids of particle volume fractions ranging from $\approx$10\% to 50\%. For the
case where no shear was applied on the ER fluid, particles were observed to form
into strings, which subsequently aggregated together. At 10\% volume fraction,
strings and small aggregates of strings are seen. At 30\% v.f. a kinetically
arrested gel is seen, where the gel is composed of hexagonal sheets of
particles. The hexagonal arrangement within the sheets is not described. The
sheets form into a labyrinthine structure similar to that seen in our
observations. The simulation remains trapped in a local potential minimum, and
is not seen to evolve further towards a crystalline ground state. In this
simulation, Brownian motion was turned off upon application of the electric
field. A similar kinetically trapped gel-like state was observed in the
simulations of Hass~\cite{Has93}, who did not observe the ER fluid to evolve
into a regular lattice. Brownian effects were neglected in this simulation.

A Brownian dynamics simulation by Tao and Jiang~\cite{Tao94} on a system
consisting of 122 particles (with volume fraction $\ge$20\%), which considers
thermal motion during structure formation, shows the ER fluid rapidly forming
chains, which aggregate into thick columns. These columns consist of
polycrystalline BCT lattice grains that are aligned along the field direction,
but misaligned in the xy direction. We observed similar column-like structures
at $\approx$25\% volume fraction. When thermal forces are neglected, the ER
fluid is described to be trapped in a local minimum energy state. However, a
sheet-like state is not reported as an intermediate state during structure
development with thermal forces included.

\bf CONCLUSION \rm\vspace{1mm}

We have verified through direct visualization that our model ER fluid reaches
the BCT structure as a ground state under most conditions, and have been able to
describe structure formation as a function of concentration. At the lowest
concentrations observed (about 10\%), BCT crystals were seen primarily to form
through chains of particles attracting each other, while a smaller fraction of
particles formed sheets that transformed into BCT structures by annealing
together. As the concentration was increased beyond 15\% the presence of chains
decreased and sheets dominated at early times, forming into complex labyrinthine
structures which lasted for hours before annealing into BCT crystals. At high
concentrations (larger than 40\%), where the initial structure was FCC, BCT
crystals were formed via defects appearing in the existing hexagonal structures.
The unusual symmetry of the BCT crystals and the martensitic FCC-BCT crystal
transition both promise applications in photonics. Unlike magnetorheological
fluids, ER fluids can be made from non light-absorbing materials. The
martensitic crystal switching we observed was on structures of size of the order
of the wavelength of light, as opposed to the FCC-BCT transition of
$\approx$45$\mu$m sized spheres described recently~\cite{She99}. The column-like
structures observed at intermediate concentrations ($\approx$25\%), were
interlinked in complex ways through strings of particles and BCT crystals, and
had wide variation in size. The structure within the columns was that of domains
of BCT which were misaligned in the plane normal to the electric field. We
observed that at a low concentration there was a sample thickness (about 10
sphere diameters) below which the preferred state was that of strings and
sheets, and BCT crystals did not form.

Data and analyzed images of our observations are available on the world wide web
at http://www.elsie.brandeis.edu.

\vspace{2mm} Acknowledgment: This work was supported by the United States
Department of Energy under Grant No. DE-FG02-94ER45522 and NSF International
Travel Grant INT-9113312 and is part of the research program of the ``stichting
voor Fundamenteel Onderzoek der Materie (FOM)'', which is financially supported
by the ``Nederlandse Organisatie voor Wetenschappelijk Onderzoek (NWO)''.

\begin{thebibliography}{100}

\bibitem{Sta83} J. E. Stangroom, {\it Phys. Technol.} {\bf 14}, 290 (1983)

\bibitem{Alf1998} A. van Blaaderen, {\it MRS Bulletin} {\bf 23}(10), 39 (1998)

\bibitem{She99} W. Wen, N. Wang, H. Ma, Z. Lin, W. Tam, C. Chan and P. Sheng {\it Phys. Rev. Lett.} {\bf 82}, 4248, (1999)

\bibitem{Pri69} J. Priestley, {\it The History and Present State of Electricity with Original
                  Experiments} 2nd ed. (London, 1769)

\bibitem{Poh78} H. A. Pohl, {\it Dielectrophoresis} p.495, (Cambridge University
   Press, 1978)

\bibitem{Win49} W. M. Winslow, {\it J. Appl. Phys.} {\bf 20}, 1137 (1949)

\bibitem{Par96} M. Parthasarathy and D. Klingenberg, {\it Materials Science and Engineering} {\bf R17}, 57 (1996)

\bibitem{Tao89} R. Tao and J. M. Sun, {\it Phys. Rev. Lett.} {\bf 67}, 398 (1991)

\bibitem{Che92} T. Chen, R. N. Zitter and R. Tao, {\it Phys. Rev. Lett.} {\bf 68}, 2555 (1992)

\bibitem{Hal90} T. C. Halsey and W. Toor, {\it Phys. Rev. Lett.} {\bf 65}, 2820 (1990)

\bibitem{Mar98} J. E. Martin, R. A. Anderson and C. P. Tigges, {\it J. Chem. Phys.} {\bf 108}, 3765 (1998)

\bibitem{Mar99} J. E. Martin, R. A. Anderson and C. P. Tigges, {\it J. Chem. Phys.} {\bf 110}, 4854 (1999)

\bibitem{Alf97} A. van Blaaderen, R. Ruel, and P. Wiltzius, {\it Nature} {\bf 385}, 321 (1997)

\bibitem{Alf95} A. van Blaaderen and P. Wiltzius, {\it Science} {\bf 270}, 1177 (1995)

\bibitem{Alf92} A. van Blaaderen and A. Vrij, {\it Langmuir} {\bf 8}, 2921 (1992)

\bibitem{Wil95} T. Wilson(ed.), {\it Confocal Microscopy}, (Academic Press, London, 1995)

\bibitem{Tao96} J. M. Sun and R. Tao, {\it Phys. Rev. E} {\bf 53}, 3732 (1996)

\newpage

\bibitem{Pus89} P. N. Pusey, W. van Megan, P. Bartlett, B. J. Ackerson, J. G. Rarity and S. M. Underwood,
{\it Phys. Rev. Lett.} {\bf 63}, 2753 (1989)

\bibitem{Ell97} M. S. Elliot, B. T. F. Bristol and W. C. K. Poon, {\it Physica A} {\bf 235}, 216 (1997)

\bibitem{Alf99} A. van Blaaderen, unpublished results

\bibitem{Law94} E. M. Lawrence, M. L. Ivey, G. A. Flores, J. Liu, J. Bibette and J. Richard,
{\it International Journal of Modern Physics B} {\bf 8}, 2765 (1994)

\bibitem{Mel93} J. R. Melrose and D. M. Heyes, {\it J. Chem. Phys.} {\bf 98}, 5873 (1993)

\bibitem{Mel92} J. R. Melrose, {\it Mol. Phys.} {\bf 76}, 635, (1992)

\bibitem{Has93} K. C. Hass, {\it Phys. Rev. E} {\bf 47}, 3362, (1993)

\bibitem{Too93} W. Toor, {\it Journal of Colloid and Interface Science} {\bf 156}, 335 {1993}

\bibitem{Mar98a} J. E. Martin, J. Odinek, T. C. Halsey and R. Kamien {\it Phys.
Rev. E} {\bf 57}(1), 756, (1998)

\bibitem{Tao94} R. Tao and Q. Jiang, {\it Phys. Rev. Lett.} {\bf 73}, 205, (1994)

\end{thebibliography}

\newpage

\begin{figure}
\centerline{\epsfig{file=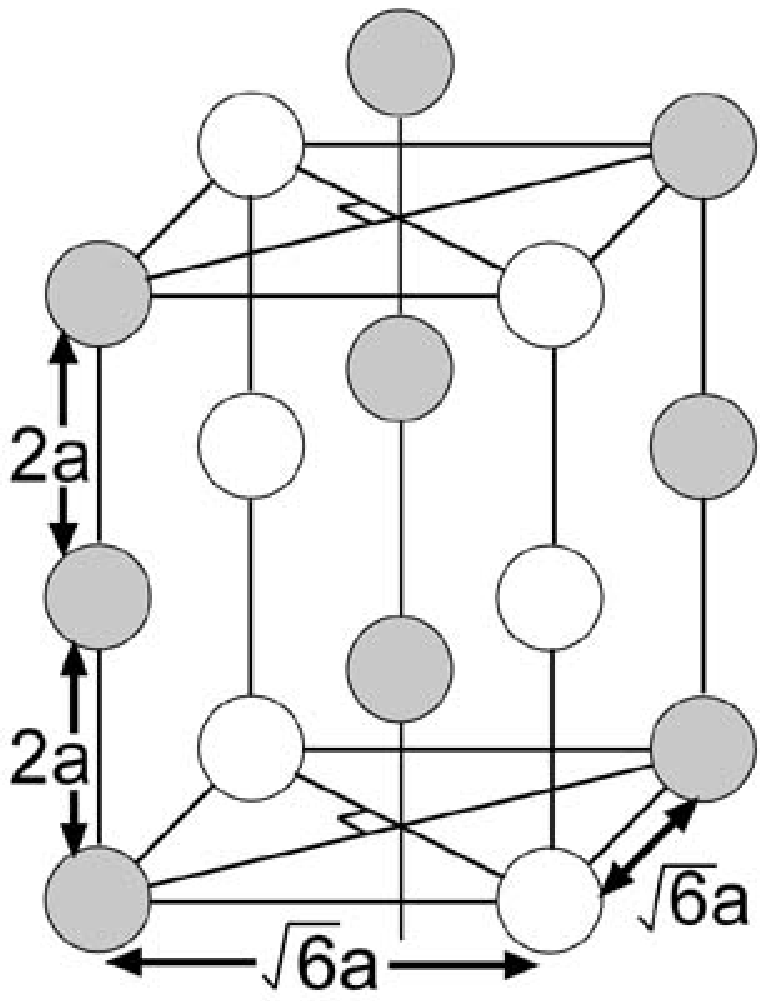,width=8.5cm}} \vspace{.25in}
\label{1}
\end{figure}

\newpage

\begin{figure}
\centerline{\epsfig{file=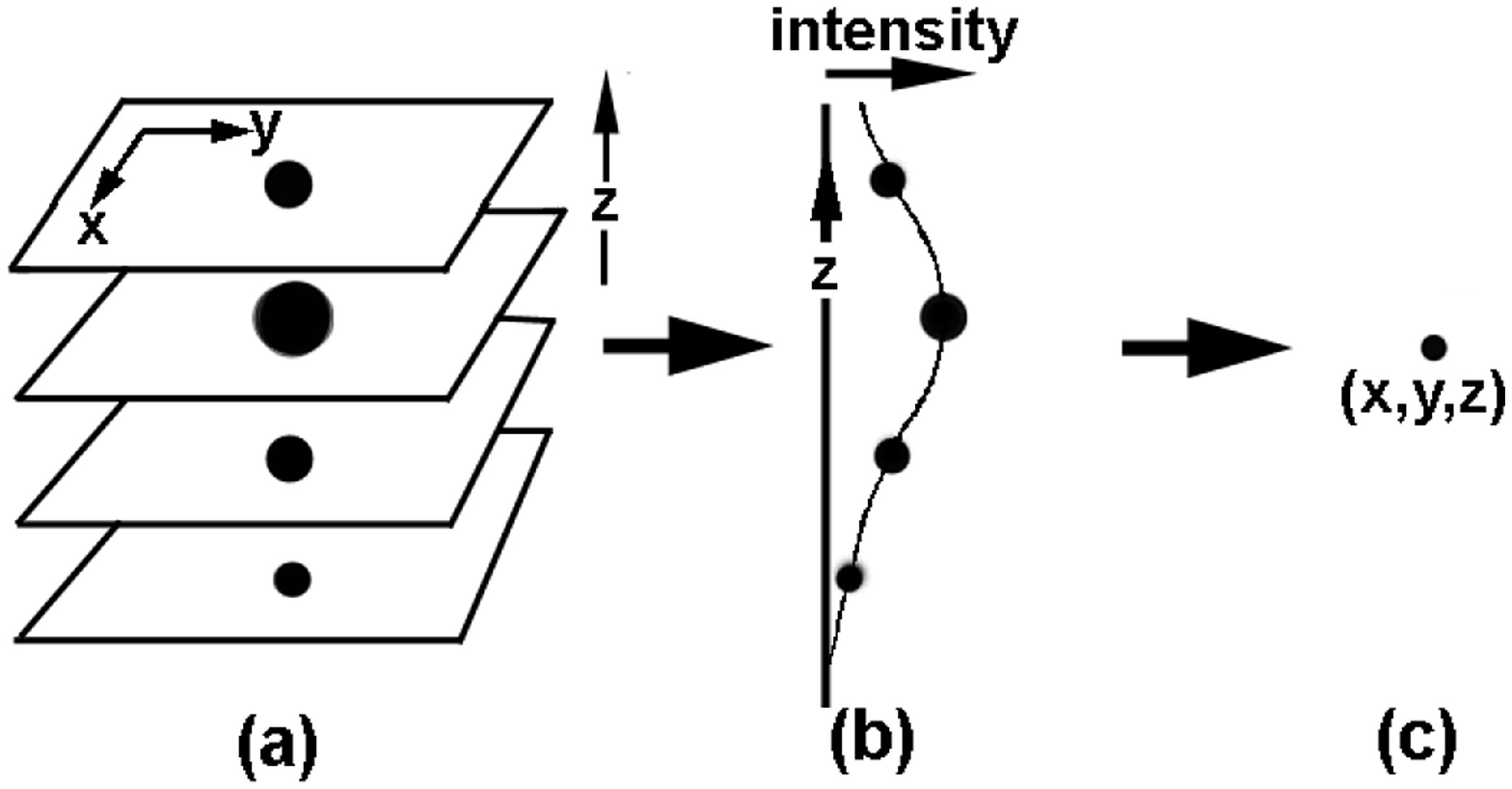,width=8.5cm}} \vspace{.25in}

\label{2}
\end{figure}

\newpage

\begin{figure}
\centerline{\epsfig{file=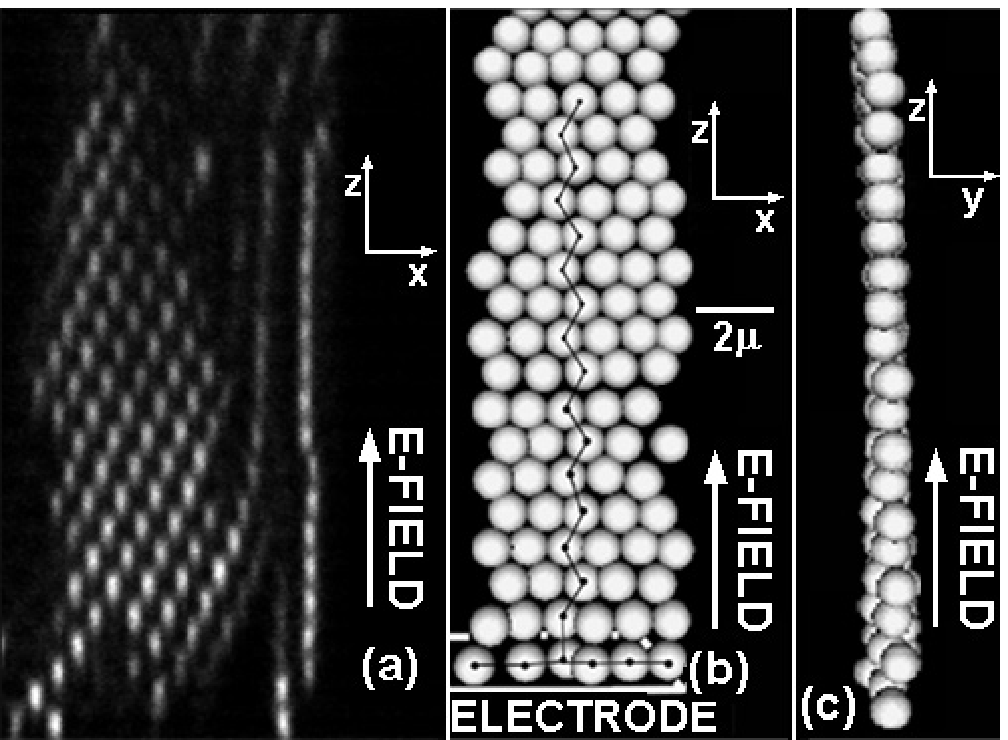,width=8.5cm}}  \vspace{.5in}
\label{3}
\end{figure}
\newpage

\begin{figure}
\centerline{\epsfig{file=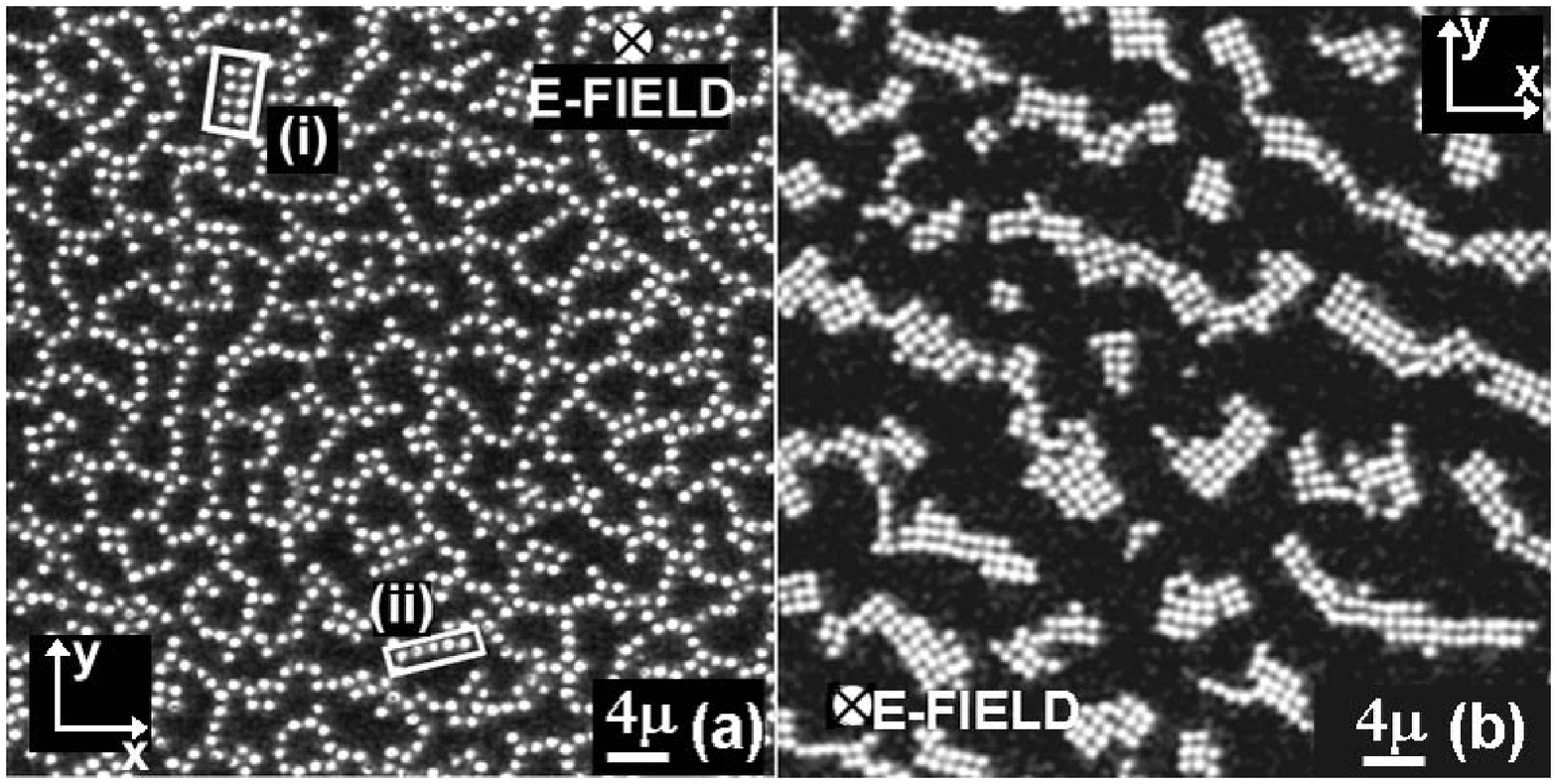,width=8.5cm}}  \vspace{.5in}
\label{4}
\end{figure}
\newpage

\begin{figure}
\centerline{\epsfig{file=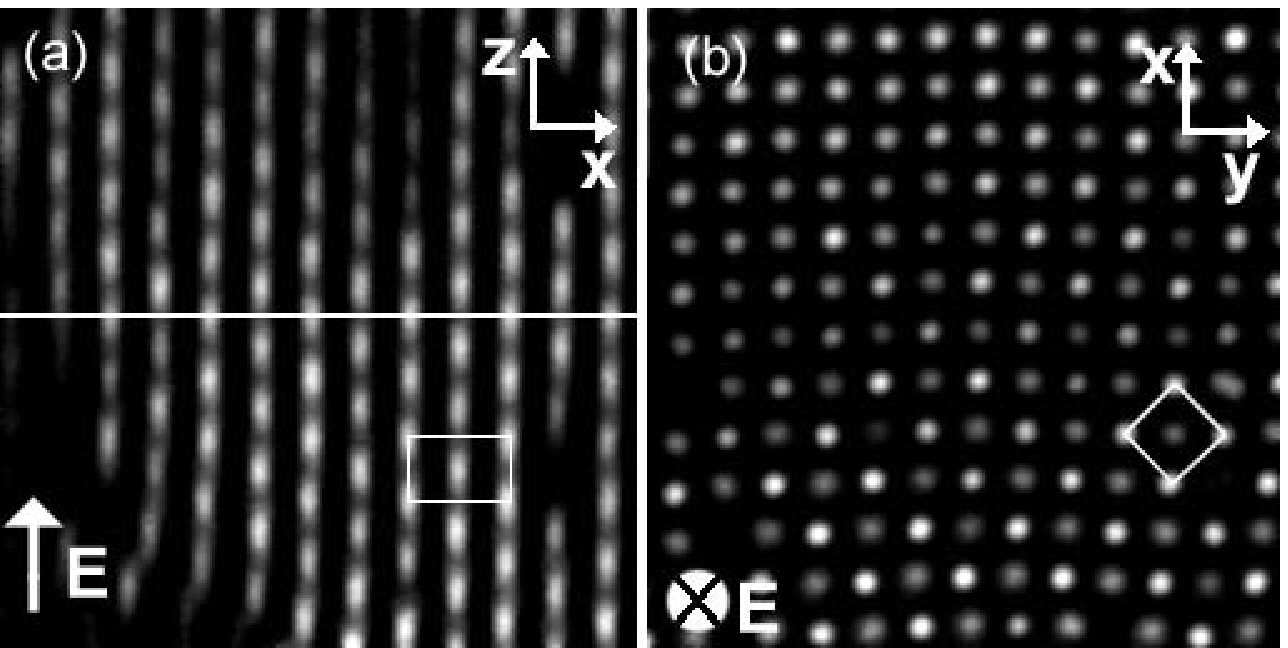,width=8.5cm}}  \vspace{.5in} \label{5}
\end{figure}

\newpage

\begin{figure}
\centerline{\epsfig{file=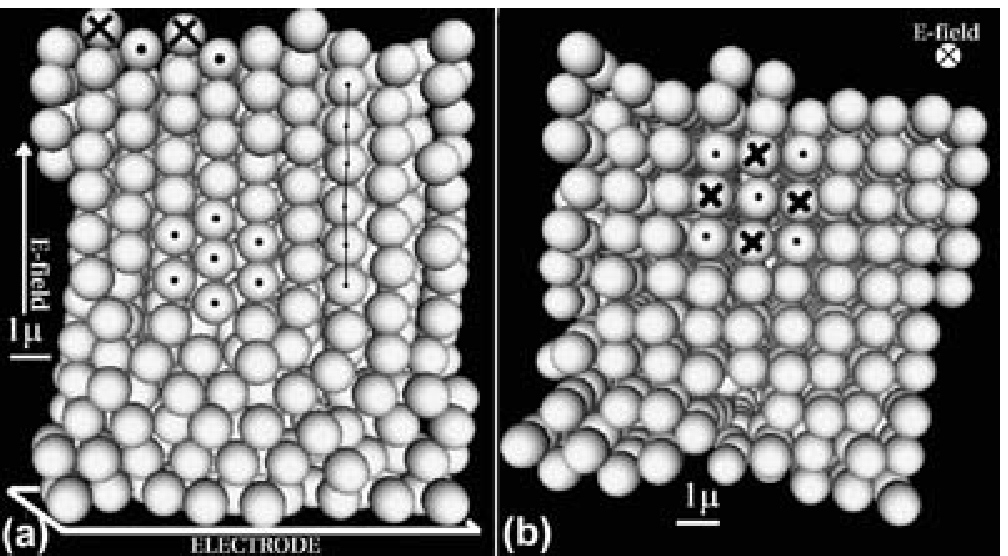,width=8.5cm}}  \vspace{.5in}
\label{6}
\end{figure}

\newpage

\begin{table}[htb] \begin{center}
\begin{tabular}{l c} \hline \hline
Structure & Energy per dipole\\ \hline Separated chains & -0.3005 \\ Cubic
lattice & -0.2618 \\ Sheet  & -0.3448 \\ BCC & -0.3401 \\ FCC &
-0.3702 \\ HCP & -0.3707 \\ BCT & -0.3813 \\ \hline \hline
\end{tabular}   \end{center}

\end{table}

\newpage

\begin{figure}
\centerline{\epsfig{file=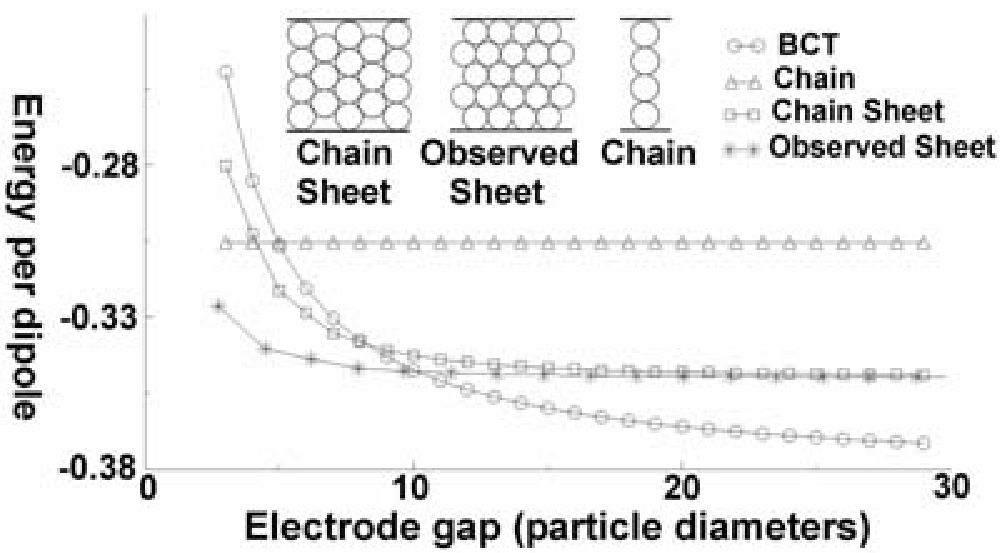,height=2.0in}} \vspace{.25in}
\label{9}
\end{figure}

\newpage

\begin{figure}
\centerline{\epsfig{file=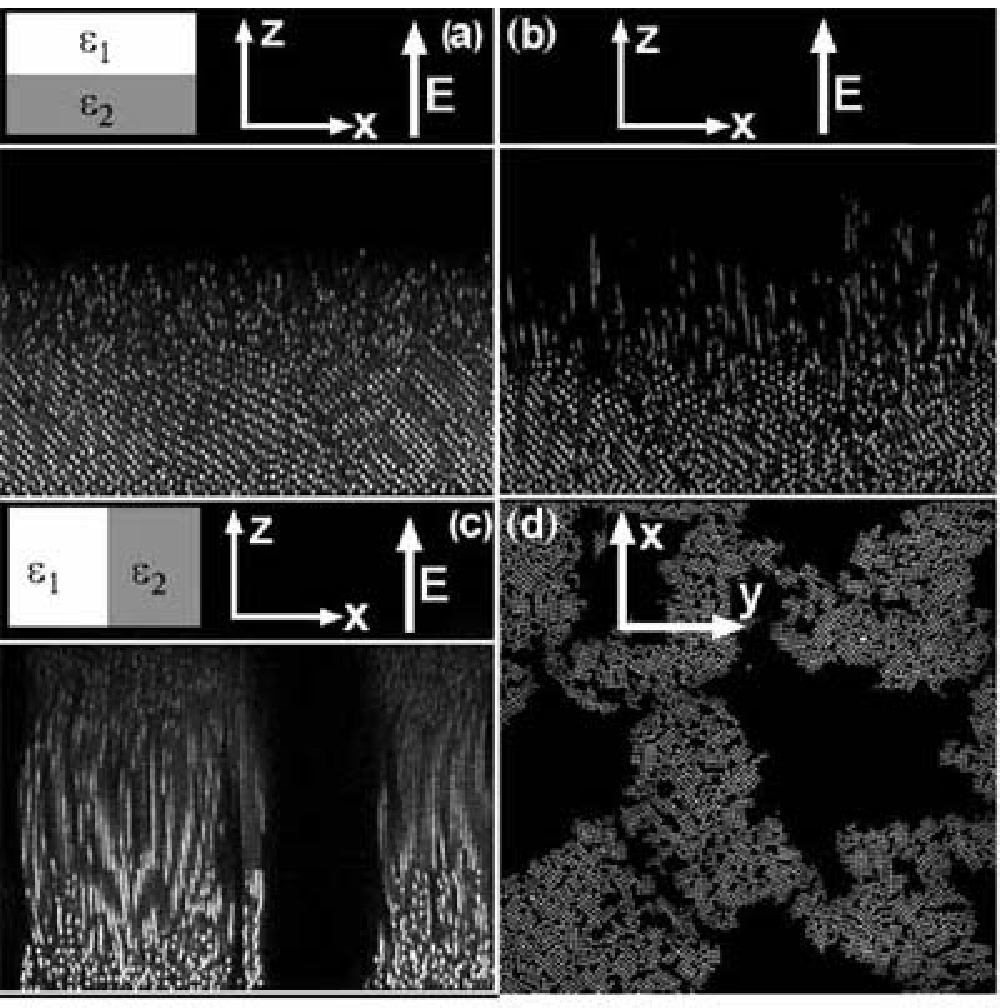,height=4.0in}} \vspace{.25in}

\label{10}
\end{figure}

\newpage

\begin{figure}
\centerline{\epsfig{file=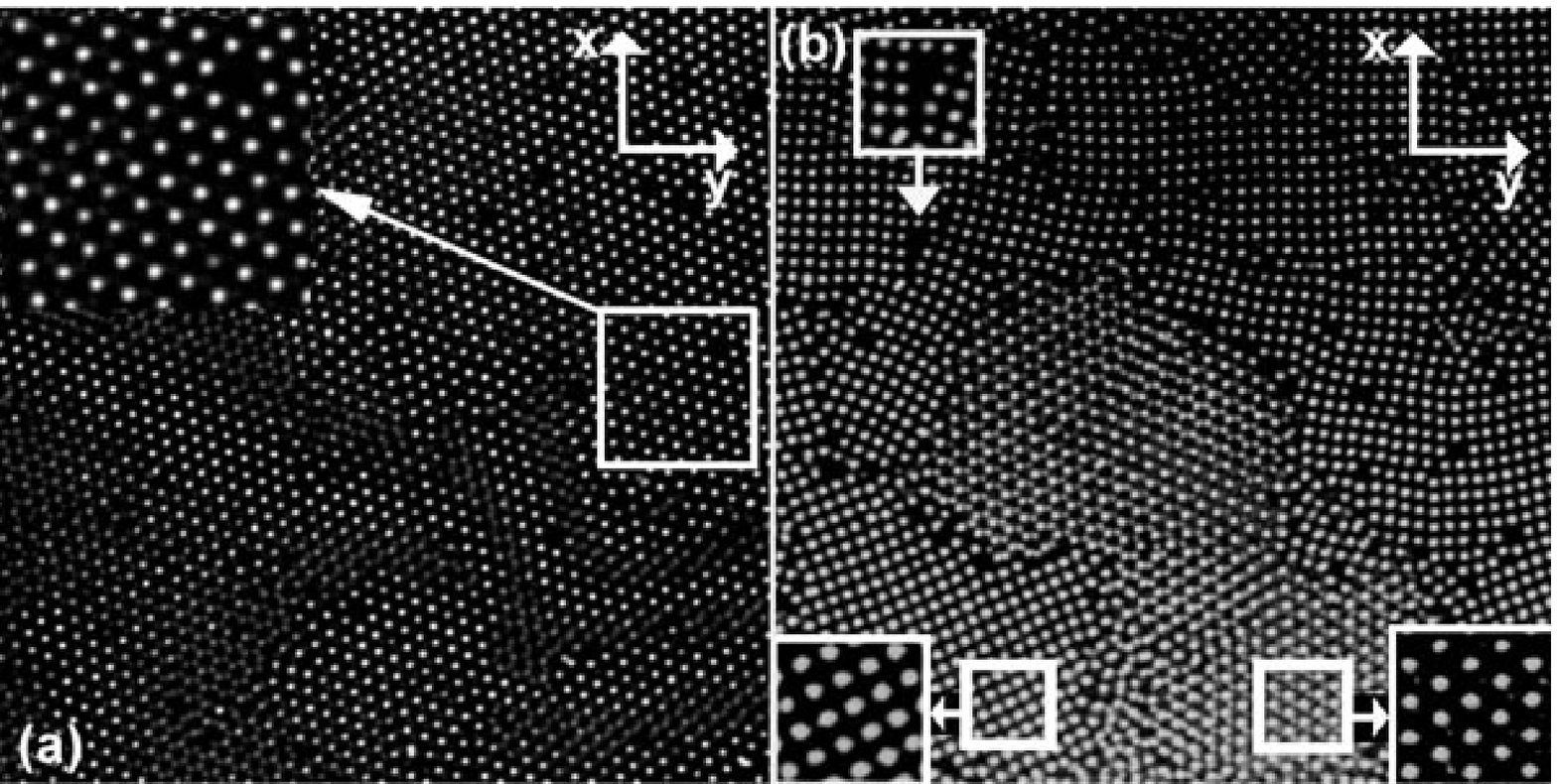,height=2.0in}} \vspace{.25in}
\label{11}
\end{figure}

\newpage
FIG. 1. The three-dimensional BCT structure. The spheres have radius $a$ (the
spheres are not drawn to scale). The crystal can be regarded as a collection of
close-packed planes. One such plane is indicated by the shaded spheres. Each
close-packed plane is a collection of chains of touching spheres, where
neighbouring chains are offset from each other by a particle radius. In the
ground state of ER fluids the chains are aligned with the electric field, thus a
section perpendicular to the electric field shows a square arrangement of
spheres (see fig.5b). The volume fraction of a BCT crystal is 0.698.

FIG. 2. Particle finding algorithm. A diagram showing the method of obtaining
particle coordinates from the confocal micrographs. The z axis is along the
optical axis and the electric field. (a)Each particle gives rise to roughly
circular regions of varying intensity in neighbouring image planes which are
separated by $\approx$0.1$\mu$m. The xy centroid coordinate of each region is
obtained by averaging coordinates weighted by intensity. (b)The regions
belonging to an individual particle are identified. (c) The final xyz coordinate
is found by an intensity-weighted average along each axis of the regions
identified in (b).

FIG. 3. Sheet structure. The field strength is 1.2 kV/mm and the field has been
on for $\approx$30 minutes. (a) x-z Confocal micrograph (raw data). There is a
sheet of particles seen in the left side of the image, with a single chain of
particles to the right of it. The bottom of the image is closest to the
objective lens of the microscope. It should be noted that since only the cores
of the silica spheres fluoresce, touching spheres are seen as separated. It is
the anisotropic point spread function that causes the spherical fluorescent
cores to appear ellipsoidal. Note also that the intensity and resolution
diminishes away from the lens due to spherical abberation.(b) View of a sheet
seen face-on. The image is taken from a rendering of reconstructed 3-d raw data
such as seen in (a). The white scale bar is 2$\mu$m. The bottom row of spheres
is touching the glass electrode (the electrode is perpendicular to the image),
and the electric field is upwards. The spheres are hexagonally close packed
within a sheet. The smallest angle between nearest neighbours and the field is
$\pm30$$^\circ$, except for the first two layers adjacent to the electrode,
where nearest neighbours are aligned along the field. (c) Side view of the same
sheet object.

FIG. 4. Labyrinth of sheets (a)View of a labyrinth of sheets looking down the
electric field. The images are raw x-y confocal micrographs. The point spread
function is symmetric in the plane (xy) perpendicular the objective. The view is
that seen after $\approx$3 minutes of the sample (volume fraction 15\%) being in
the E-field. The sheets form within seconds of the E-field being turned on, and
evolve into the BCT structure over hours, indicated by (i) in the image above.
The structure of a sheet such as indicated by (ii) in the image is shown in
figures 3b and 3c. The white bar is 4$\mu$m. Note that the raw data only shows
the (fluorescent) cores of the spheres, making it possible to distinguish
touching spheres. Spheres within $\pm0.5$$\mu$m of the image plane contribute to
the image. The image is from a plane 20$\mu$m from an electrode in a sample with
an electrode gap of 70$\mu$m. (b) After several hours, the sheets such as seen
in (a) anneal together and form collections of BCT structures. The field is
perpendicular to the image. Each BCT cluster extends long distances in the field
direction.

FIG. 5. Body Centered Tetragonal crystal. Raw 3D data set of a BCT crystal
consisting of $13\mu\mbox{m} \times 13\mu\mbox{m}$ x-y planes separated in the
z-direction by 0.09$\mu$m. The image (a) is a digital interpolation of data from
a sequence of x-y planes such as seen in (b). (a) shows a view along a plane
parallel to the E-field, showing the centered rectangular lattice of dimension
2$\sqrt3 a \times 2a$ (110 plane) of the BCT lattice, where the spheres have
radius $a$. In this plane the BCT structure consists of chains of spheres
aligned along the field direction. Neighbouring chains are offset along the
field by one particle radius. (b) shows a view looking down the E-field showing
the square $\sqrt6 a
\times
\sqrt6 a$ lattice (001 plane) of the BCT lattice. The plane (b) is a section
orthogonal to (a) such as along the line indicated in (a). The alternating
intensity pattern seen in (b) arises because adjacent chains of spheres are
offset in and out of the plane by one particle radius.

FIG. 6. BCT Crystal. Two views of a BCT crystal rendered in 3D after obtaining
 coordinates of the centers of each particle from raw data such as seen in
Figure 5. The view is of a portion of a larger crystalline region. (a) shows a
view looking in a plane parallel to the E-field. The particles connected with a
line show a chain parallel to the E-field. The view shows the hexagonal 110
plane of the BCT crystal. As in (b), the x and $\bullet$ show neighbouring
chains of particles that are out of register by a particle radius. The spheres
in the first few layers adjacent to the electrode are arranged in random stacked
hexagonal planes parallel to the electrode. (b) shows a view perpendicular to
the E-field. The $\bullet$ and x symbols represent chains of particles aligned
with the E-field, and chains marked with a $\bullet$ are out of register with
the x chains by one particle radius.

Table 1: Dipolar energy per particle for various infinite sized lattices. Energy
is in units of
\begin{math}p^2/a^3\epsilon_f \end{math}, where the dipole moment \begin{math}
p=a^3\epsilon_fE(\epsilon_p-\epsilon_f)/(\epsilon_p+2\epsilon_f). \end{math}
\begin{math}\epsilon_p\end{math} and \begin{math}\epsilon_f\end{math} are the
dielectric constants of the particle and solvent respectively, and E is the
magnitude of the electric field.

FIG. 7. Dipole energy per particle vs. electrode gap. Image charges are included
in this calculation. The energy per particle is in units of
$p^2$$/$$a^3$$\epsilon_f$, where
 $zp=a^3$$\epsilon_f$E($\epsilon_p$-$\epsilon_f$)/($\epsilon_p$+2$\epsilon_f$).
 $\epsilon_p$ and $\epsilon_f$ are the dielectric constants of the particle and
 solvent respectively, and E is the magnitude of the electric field. `Observed
 sheets' are sheets with the configuration observed in the experiments, while
 `chain sheets' are close-packed planar sheets consisting of chains of spheres
 aligned along the electric field, where neighbouring chains are offset along
 the electric field direction by one particle radius. At an electrode spacings
 of less than 10 particle diameters the observed sheets have a lower energy than
 the BCT crystal.

FIG. 8. Field response of sedimented colloid. (a) At a particle volume fraction
of $\approx$25\% the colloid is left to sediment to the bottom electrode before
a E-field is applied. The horizontal white lines represent the transparent
electrodes. (b) Shows the sedimented sample as shown in (a) transforming into a
columnlike structure upon application of a E-field. The image is taken 2 minutes
after the field was on. Chains of particles are seen to form along the field.
(c) After a few hours the spheres arrange into cross-linked columns formed
parallel to the field. (d) A view of the columns from a plane perpendicular to
the field. The square-like arrangement of the particles within the columns
indicates that the structure within the columns is BCT. If the sphere-rich
region and solvent are treated as dielectric fluids with different dielectric
constants placed between the plates of a parallel plate capacitor, then the
configuration seen in (c) (and shown schematically in the upper left corner of
(c)), where the fluids are separated into regions with their interfaces parallel
to the field, has lower electrostatic energy than the configuration in (a),
where the fluids are separated in the sedimented state. This explains the
tendency to form columns.

FIG. 9. Field induced solid-solid transition. The image shows raw confocal
microscope data of a sample of volume fraction $\approx$45\%. (a) Shows a plane
parallel to the electrodes before the E-field was turned on. The spheres are
arranged as hexagonal planes with many defects, parallel to the electrodes. The
structure is FCC. The image is from a plane 20$\mu$m from an electrode, and the
electrode gap is 80$\mu$m. (b) Shows the same area about 6 hours after an
E-field is applied (perpendicular to the image plane). Large areas of the
crystal have transformed into BCT order, identified by the square
configurations.

\end{document}